\documentclass[pra,twocolumn,aps]{revtex4}

\usepackage[english]{babel}
\usepackage[utf8]{inputenc}
\usepackage{graphicx}
\usepackage{siunitx}
\usepackage{amsmath}
\usepackage{amsfonts}
\usepackage{amssymb}
\usepackage{epsfig}
\usepackage{graphicx}

\newcommand{\beq}{\begin{equation}}
\newcommand{\eeq}{\end{equation}}
\newcommand{\beqa}{\begin{eqnarray}}
\newcommand{\eeqa}{\end{eqnarray}}

\begin{document}

\title{First and second sound in two-dimensional bosonic 
and fermionic superfluids}

\author{L. Salasnich$^{1,2,3}$, A. Cappellaro$^{4}$, K. Furutani$^{1,3}$, 
A. Tononi$^{5}$, G. Bighin$^{6}$}

\address{$^{1}$Dipartimento di Fisica e Astronomia 
``Galileo Galilei'' and Padua QTech, 
Universit\`a di Padova, Via Marzolo 8, 35131 Padova, Italy \\
$^{2}$Istituto Nazionale di Ottica (INO) del Consiglio Nazionale 
delle Ricerche (CNR), Via Nello Carrara 1, 50019 Sesto Fiorentino, Italy\\
$^{3}$Istituto Nazionale di Fisica Nucleare (INFN), Sezione di Padova, 
Via Marzolo 8, 35131 Padova, Italy\\
$^{4}$Institute of Science and Technology Austria (ISTA), 
Am Campus 1, 3400 Klosterneuburg, Austria \\
$^{5}$Universit\'e Paris-Saclay, CNRS, LPTMS, 91405 Orsay, France\\
$^{6}$Institut fur Theoretische Physik, Universitat Heidelberg, 
Philosophenweg 19, 69120 Heidelberg, Germany}


\begin{abstract}
{We review our theoretical results about the sound propagation 
in two-dimensional (2D) systems of ultracold fermionic and bosonic atoms.  
In the superfluid phase, characterized by the spontaneous symmetry breaking 
of the $U(1)$ symmetry, there is the coexistence of first and 
second sound. In the case of weakly-interacting { repulsive} bosons, 
we model 
the recent measurements of the sound velocities of $^{39}$K atoms 
in 2D obtained in the weakly-interacting regime and 
around the Berezinskii-Kosterlitz-Thouless (BKT) superfluid-to-normal 
transition temperature. In particular, we perform a quite accurate 
computation of the superfluid density and show that it is reasonably 
consistent with the experiment. For superfluid { attractive} fermions, 
we calculate the first and second sound velocities across 
the whole BCS-BEC crossover. In the low-temperature 
regime we reproduce the recent measurements of first-sound speed 
with $^6$Li atoms. We also predict that only in the finite-temperature 
BEC regime there is mixing between sound modes.}
\end{abstract}

\maketitle

\section{Introduction}

In this review paper the propagation of first and 
second sound in two-dimensional (2D) systems of ultracold fermionic and 
bosonic atoms is examined in light of our current theoretical 
findings. As well known, the second sound exists only 
in the $U(1)$ symmetry-broken superfluid phase. 
We discuss a quite accurate determination of the superfluid density 
in the case of weakly-interacting { repulsive} bosons, 
finding a good agreement 
with the experiment to model the recent measurements \cite{chris2021} 
of the sound 
velocities of $^{39}$K atoms in 2D obtained in the weakly-interacting regime 
and around the Berezinskii-Kosterlitz-Thouless (BKT) 
superfluid-to-normal transition temperature \cite{furutani2021}. 
We also analyze the first and second sound 
velocities across the whole BCS-BEC crossover for superfluid { 
attractive} fermions. By considering $^6$Li atoms we simulate and analyze 
the most recent measurements 
\cite{bohlen2020} of the first sound velocity in the low-temperature regime. 
This velocity is the only one triggered by a density probe because 
the decoupling of density and entropy fluctuations makes it possible 
\cite{tononi2021}. The main results discussed here have been presented 
at the International Workshop ``Quantum Transport with ultracold atoms'' 
(Dresden, 2022). 

According to Landau's two fluid theory \cite{landau1941}
of superfluids the total number density $n$ of a 
system in the superfluid phase can be written as 
\beq 
n = n_\text{s} + n_\text{n} \; , 
\eeq
where $n_\text{s}$ is the superfluid density and $n_\text{n}$ is the normal 
density. At the critical temperature $T_c$ one has $n_{n}=n$ and, 
correspondingly, $n_\text{s}=0$. 
Following Landau, in a superfluid a local perturbation excites two 
wave-like modes - first and second sound - which propagate 
with velocities $u_1$ and $u_2$. 
These velocities are determined by the positive solutions 
of the algebraic biquadratic equation 
\beq 
u^4 - (c_{10}^2+c_{20}^2) u^2 + c_T^2 c_{20}^2 = 0 \; .   
\label{biquad}
\eeq
The first sound $u_1$ is the largest of the two positive roots 
of Eq.~(\ref{biquad}) while the second sound $u_2$ is the 
smallest positive one.
In the biquadratic equation (\ref{biquad}) there is the adiabatic 
sound velocity
\beq 
c_{10} = \sqrt{{1\over m} \left({\partial P 
\over \partial n}\right)_{\bar{S},V} } \; 
\label{c10}
\eeq
with $\bar{S}=S/N$ the entropy per particle, $V=L^2$ 
the 2D volume (area) of a square of size $L$, and $N$ the total number of 
identical particles. There is also the entropic sound (or Landau) velocity  
\beq 
c_{20} = \sqrt{{1\over m} {{\bar S}^2 \over 
\left({\partial {\bar S}\over \partial T}\right)_{N,V}} 
{n_\text{s}\over n_n} } \;  
\label{c20}
\eeq
with $n_s/n_n$ the ratio between superfluid and normal density 
and $m$ the mass of each particle, and the isothermal sound velocity 
\beq
c_T = \sqrt{{1\over m} \left({\partial P 
\over \partial n}\right)_{T,V} } \;  
\label{cT}
\eeq
with $P$ the pressure and $T$ the temperature. 
Thus, having the equation of state and the superfluid fraction 
of the system under investigation one can determine the first sound 
velocity $u_1$ and the second sound velocity $u_2$ by solving 
Eq. (\ref{biquad}), namely  
\beq 
u_{1,2} = \sqrt{{1\over 2}  (c_{10}^2+c_{20}^2) \pm 
{1\over 2} \sqrt{ (c_{10}^2+c_{20}^2)^2 - 4 c_T^2 c_{20}^2}} \; . 
\label{boh}
\eeq

\section{Weakly-interacting 2D Bose gas}

The Helmholtz free energy \cite{tononi2021} of a weakly-interacting 
two-dimensional gas of { purely repulsive} identical bosons of mass $m$ 
can be written as ($\hbar=k_{\mathrm{B}}=1$) 
\beq
F=F_{0}+F_{\mathrm{Q}}+F_{T} =
\frac{g}{2}\frac{N^{2}}{L^{2}}+\frac{1}{2}\sum_{\bf{p}}E_{p}
+T\sum_{\bf p}\ln{\left[ 1 - e^{-E_{p}/T}\right]} \; ,
\label{freeenergy}
\eeq 
where $F_{0}$ is the mean-field zero-temperature free energy 
with $g{ >0}$ is the Bose-Bose interaction strength.  
$F_{T}$ is the low-temperature free energy and
\beq
E_{p}=\sqrt{\frac{p^{2}}{2m}\left(\frac{p^{2}}{2m}+2gn\right)} \; ,
\eeq
is the familiar Bogoliubov spectrum of elementary excitations. 
For the sake of completeness, we emphasize some formal analogy \cite{sala-power}
between this Bogoliubov spectrum of bosonic particles, which can be 
derived from the Gross-Pitaevskii equation \cite{gross1961,pitaevskii1961}, 
and the neural spectrum which can be deduced from the Amari equation of the 
brain \cite{amari1977}. { Indeed, the neural field equation of Amari  
resembles an imaginary-time Gross-Pitaevskii equation with 
a nonlocal term. The elementary (linearized) excitations of the Amari equation 
around a uniform configuration are the analog 
of the Bologlibov spectrum of the Gross-Pitaevskii equation.}  
The quantum correction $F_{\mathrm{Q}}$ in the free energy is obviously 
ultraviolet divergent and requires a regularization procedure. 
Dimensional regularization \cite{sala2016} leads to 
\beq 
F_{\mathrm{Q}}= -L^2 \frac{m}{8\pi} \left[ 
\ln{\left({\epsilon_{\Lambda}\over g n} 
\right)} - \frac{2}{\eta} \right] \, \left(g n\right)^2 \; , 
\eeq
where $\epsilon_{\Lambda}=4e^{-2\gamma-1/2}/\left(ma_{\mathrm{2D}}^{2}\right)$ 
is a cutoff energy, $\gamma=0.577$ is the Euler-Mascheroni 
constant, $a_{\mathrm{2D}}$ is the 2D $s$-wave scattering length, and 
\beq 
\eta = {m g\over 2\pi}  
\eeq
is the adimensional gas parameter \cite{furutani2021}. Moreover, one also finds 
\beq
\frac{\epsilon_{\Lambda}}{gn}=\frac{2\pi }{N}
\frac{e^{-2\gamma-1/2+2/\eta}}{\eta} \; .
\label{cutoff}
\eeq

All the thermodynamic quantities can be obtained from the 
Helmholtz free energy of Eq.~(\ref{freeenergy}). For instance, 
the pressure $P$ is given by 
\beq
P=-\left(\frac{\partial F}{\partial L^{2}}\right)_{N,T} 
\eeq
while the the entropy reads 
\beq
S=\left(\frac{\partial}{\partial T}\frac{F}{N}\right)_{N,L^2} \; . 
\eeq

Instead, the normal density $n_n$ can be extracted from the Landau formula 
\beq
n_n=- \int\frac{d^{2}{\bf p}}{(2\pi)^{2}} \, 
{p^2\over 2m}\frac{df_{\mathrm{B}}(E_{p})}{dE_{p}} \; ,
\label{rhon}
\eeq
where $f_{\mathrm{B}}(E)=1/\left(e^{E/T}-1\right)$ is the 
Bose-Einstein distribution. 

\begin{figure}[t]
\centerline{\epsfig{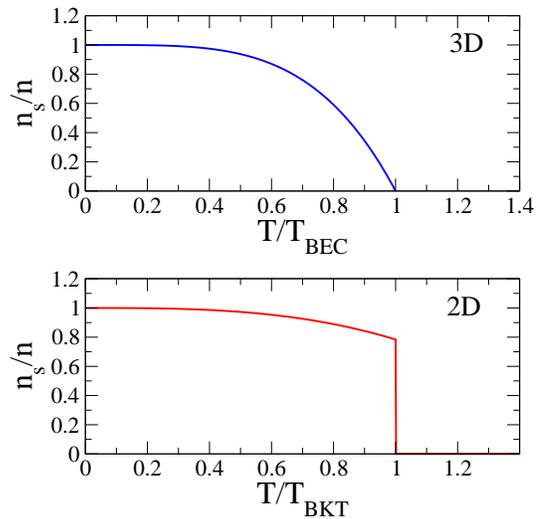}}
\caption{Typical behavior of superfluidy fraction $n_s/n$ 
vs adimensional temperature $T/T_c$ 
in three-dimensional (3D) and two-dimensional (2D) superfluid systems. 
Notice that only in the 2D case there is a jump of the superfluid fraction 
at the Berezinskii-Kosterlitz-Thouless critical temperature $T_{BKT}$ 
of the superfluid-to-normal phase trasition.} 
\label{fig1}
\end{figure} 

Actually, the Landau formula for the normal density does not 
take into account the formation of quantized vortices 
and anti-vortices by increasing the temperature. 
These quantized vortices are crucial 
for the 2D Bose gas to obtain the phenomenology predicted by 
Berezinskii \cite{ber1972} and Kosterlitz-Thouless \cite{kos1972}. 
The presence of quantized vortices renormalize the superfluid 
density $n_s=n-n_n$. The renormalized superfluid density    
$n_{s}(t=+\infty)$ is obtained by solving the Nelson-Kosterlitz
renormalization group equations \cite{nelson1977}
\begin{eqnarray}
\nonumber
\partial_t \, K^{-1}(t) &=& 4 \pi^3 y^2(t) \\
\partial_t \, y(t) &=& [2-\pi K(t)] \, y(t)
\label{rg}
\end{eqnarray}
where $K(t)=n_{\mathrm{s}}(t)/T$, with $n_{\mathrm{s}}(t)$ the 
superfluid density at the adimensional fictitious time $t$, and 
$y(t)=\exp\left[-\mu_{\mathrm{c}}(t)/T\right]$ is the fugacity, 
where $\mu_{\mathrm{c}}(t)$ is the vortex chemical potential at 
fictitious time $t$. { In particular, the initial superfluid 
density $n_{\mathrm{s}}(0)$ of the flow is the one obtained from $n_n$ of 
Eq. (\ref{rhon}) as $n_{\mathrm{s}}(0)=n-n_n$. The initial vortex 
chemical potential is instead given by the expression 
$\mu_{\mathrm{s}}(0)=\pi^2n_s(0)/(2m)$. We emphasize that 
in the determination of pressure and entropy 
one should take into account also the vortex contribution.  
However, in order to make the theoretical scheme more tractable 
we have included the quantized vortices only for the 
renormalized superfluid density. Another relevant 
issue is the fact in the experiments the superfluids have a finite size. 
To describe consistently the finite size of the system, 
we solve Eqs. (\ref{rg}) up to a maximum value 
$t_{max} = ln(A^{1/2}/\xi)$ of the adimensional 
fictitious time $t$, where $A$ is the area of the system 
and $\xi$ is the healing length, 
which is practically the size of the vortex core.} 

For 3D superfluids the transition to the normal state is a 
BEC phase transition, while in 2D superfluids the transition to the 
normal state is something different: a topological phase transition. 
An important prediction of the Kosterlitz-Thouless transition is that, 
contrary to the 3D case, in 2D the superfluid fraction  
$n_s/n$ jumps to zero above the Berezinskii-Kosterlitz-Thouless 
critical temperature $T_{BKT}$. See Fig. \ref{fig1}. 

\begin{figure}[t]
\centerline{\epsfig{file=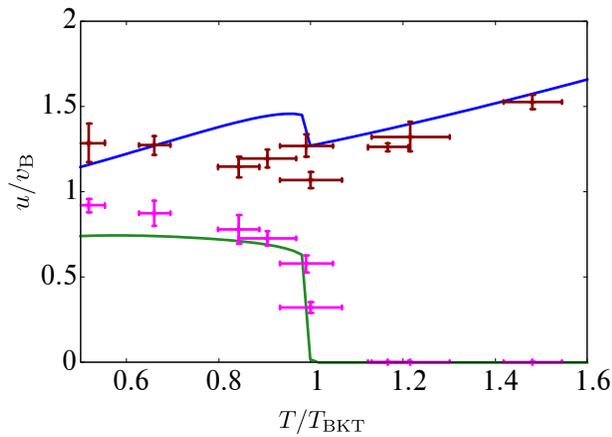,width=8cm,clip=}}
\caption{Sound velocities vs adimensional temperature. 
Here $v_B=gn$ is the Bogoliubov velocity, $N = 2178$ is the number 
of atoms and $\eta=0.102$. The blue line is our first sound 
velocity $u_1$ while the green line is our second sound velocity $u_2$. 
The dots with error bars are the experimental data obtained by 
Christodoulou {\it et al.} \cite{chris2021}. 
Figure adapted from Ref. \cite{furutani2021}.} 
\label{fig2}
\end{figure}

In Fig. \ref{fig2}, we report our theoretical first and second sound 
velocities as a function of the adimensional temperature  
in comparison with recent experimental data near $T_{BKT}$ \cite{chris2021}. 
As shown by the figure, the agreement between our theory and 
the experimental results is quite good. 

{ Another relevant phenomenon is the hybridization with quasi-crossing 
of first sound $u_1$ and second sound $u_2$ which appears at a 
characteristic temperature $T_{hyb}$. 
In particular, when the hybridization temperature $T_{hyb}$ 
is crossed, there is an 
inversion of the role of density and entropy oscillations in the propagation 
of sounds. As discussed in detail in Ref. \cite{furutani2021}, 
a density perturbation excites mainly $u_1$ below the 
$T_{hyb}$ and instead probes mainly $u_2$ 
above $T_{hyb}$. The opposite happens by imposing a temperature gradient 
in the superfluid. Numerically, we have found that $T_{hyb}$ grows 
by increasing the repulsive Bose-Bose interaction strength 
and eventually $T_{hyb}$ coincides with $T_{BKT}$ \cite{furutani2021}.}

\section{2D Fermi gas in the BCS-BEC crossover}

In 2004 the 3D BCS-BEC crossover has been 
observed with ultracold gases made of two-component { attractive} 
fermionic $^{40}$K or $^6$Li atoms \cite{regal2004,zwierlein2004,kinast2004}. 
This crossover is obtained using a Fano-Feshbach resonance to change 
the 3D s-wave scattering length $a_F$ of the 
inter-atomic potential. More recently also the 2D BEC-BEC crossover has been 
achieved experimentally \cite{makhalov2014,ries2015} 
with a Fermi gas of two-component $^6$Li atoms. 

Two dimensional realistic interatomic attractive potentials 
always have a bound state, 
in contrast to the 3D situation. In particular \cite{mora2003}, 
the binding energy $\epsilon_B>0$ of two fermions is related 
to the 2D scattering length $a_F$ by 
\beq 
\epsilon_B= {4\over e^{2\gamma}}{1\over m a_F^2} \; ,   
\label{eb-af}
\eeq
where $\gamma=0.577$ is the Euler-Mascheroni constant. Moreover, 
the attractive interaction of strength $g>0$ 
of s-wave pairing is related to the binding energy by 
the expression \cite{randeria1989} 
\beq 
\frac{1}{g} 
= \frac{1}{2L^2} \sum_{\bf k} \frac{1}{{k^2\over 2m} + 
\frac{1}{2} \epsilon_B} \; . 
\label{g-eb}
\eeq 

To study the 2D BCS-BEC crossover we adopt the formalism of functional 
integration \cite{nagaosa1999}. 
The partition function ${\cal Z}$ of the uniform 
system with fermionic fields $\psi_{s} ({\bf r},\tau )$ 
at temperature $T$, in a $2$-dimensional volume $V=L^2$, 
and with chemical potential $\mu$ reads
\beq 
{\cal Z} = \int {\cal D}[\psi_{s},\bar{\psi}_{s}] 
\ \exp{\left\{ -{S} \right\} } \; , 
\eeq
where $\beta \equiv 1/T$ and 
\beq 
S = \int_0^{\beta} 
d\tau \int_{L^2} d^2{\bf r} \ {\cal L}
\eeq
is the Euclidean action functional with Lagrangian density
\beq 
{\cal L} = \bar{\psi}_{s} \left[ \partial_{\tau} 
- \frac{1}{2m}\nabla^2 - \mu \right] \psi_{s} 
- g \, \bar{\psi}_{\uparrow} \, \bar{\psi}_{\downarrow} 
\, \psi_{\downarrow} \, \psi_{\uparrow} 
\eeq
where $g>0$ is the strength of the attractive 
the s-wave coupling between fermions with opposite spin. 

In particular, we are interested in the grand potential $\Omega$, 
given by 
\beq 
\Omega = - {1\over \beta} \ln{\left( {\cal Z} \right)} \simeq - {1\over \beta} 
\ln{\left( {\cal Z}_{mf} {\cal Z}_g \right)} = \Omega_{mf} + \Omega_{g} 
\; , 
\eeq 
where 
\beq 
{\cal Z}_{mf} = \int {\cal D}[\psi_{s},\bar{\psi}_{s}]\, 
\exp{\left\{ - {S_e(\psi_s, \bar{\psi_s}, 
\Delta_0)} \right\}} \;  
\eeq
is the mean-field partition function and 
\beq
{\cal Z}_g = \int {\cal D}[\psi_{s},\bar{\psi}_{s}]\, 
{\cal D}[\eta,\bar{\eta}] \ 
\exp{\left\{ - {S_g(\psi_s, \bar{\psi_s},
\eta,\bar{\eta},\Delta_0)} \right\}} 
\eeq
is the partition function of Gaussian pairing fluctuations. 

After functional integration over quadratic fields, 
one finds that the mean-field grand potential 
reads \cite{altland2006} 
\beq
\Omega_{mf} = {\Delta_0^2\over g} L^2 
+ \sum_{\bf k} \left( {k^2\over 2m} - \mu - E_{sp}({\bf k}) 
- {2\over \beta } \ln{(1+e^{-\beta\, E_{sp}({\bf k})})} 
\right) 
\; 
\eeq
where 
\beq 
E_{sp}({\bf k}) = 
\sqrt{\left({k^2\over 2m}-\mu\right)^2+\Delta_0^2}
\eeq
is the spectrum of fermionic single-particle excitations. 

The Gaussian grand potential is instead given by 
\beq
\Omega_{g} = {1\over 2\beta} \sum_{Q} \ln{\mbox{det}({\bf M}(Q))} \; ,  
\label{goduria}
\eeq
where ${\bf M}(Q)$ is the inverse propagator of Gaussian 
fluctuations of pairs and $Q=({\bf q},i\Omega_{m})$ is the 4D 
wavevector with $\Omega_{m}=2\pi m/\beta$ the Matsubara frequencies 
and ${\bf q}$ the 3D wavevector \cite{diener2008}. 

\begin{figure}[t]
\centerline{\epsfig{file=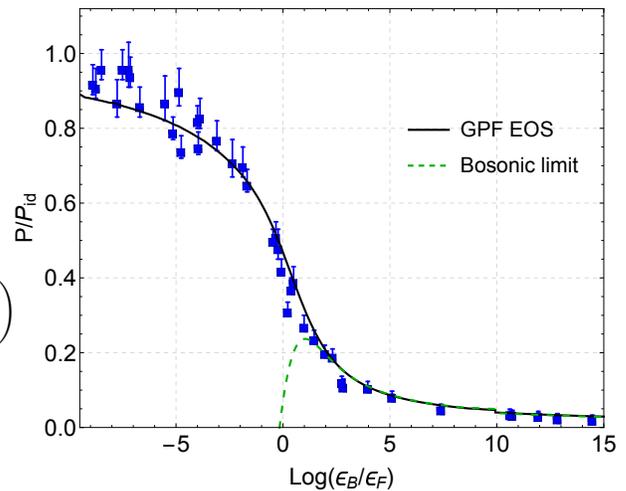,width=8cm,clip=}}
\caption{Zero-temperature scaled pressure $P/P_{id}$ vs scaled binding 
energy $\epsilon_B/\epsilon_F$. Filled squares with error bars: 
experimental data of Makhalov {\it et al.} \cite{makhalov2014}. 
Solid line: our regularized Gaussian pair fuctuation (GPF) 
theory. Figure adapted from Ref. \cite{bighin2018}.} 
\label{fig3}
\end{figure}

The sum over Matsubara frequencies is quite complicated and it 
does not give a simple expression. An approximate 
formula \cite{taylor2006} is 
\beq
\Omega_{g} \simeq {1\over 2} 
\sum_{\bf q} E_{col}({\bf q}) + {1\over \beta }
\sum_{\bf q} \ln{(1- e^{-\beta\, E_{col}({\bf q})})} \; ,   
\eeq
where 
\beq 
E_{col}({\bf q}) = \omega({\bf q}) 
\eeq
is the spectrum of bosonic collective excitations with $\omega({\bf q})$ 
derived from 
\beq 
\mbox{det}({\bf M}({\bf q},\omega)) = 0 \; . 
\eeq
{ It is important to stress that the zero-point energy of the collective 
excitations is divergent. However, by using the convergence factor 
renormalization procedure (see Ref. \cite{sala2016} 
for a review of renormalization methods for the zero-point energy 
of ultracold atoms) one extracts a reliable finite contribution.}  
In Fig. \ref{fig3} we plot the pressure $P=-\Omega/L^2$ of the 
2D Fermi gas in the BCS-BEC crossover comparing our zero-temperature 
theoretical results \cite{bighin2016} with the available 
experimental data \cite{makhalov2014}. The agreement between theory 
and experiment is extremely good only including Gaussian fluctuations. 
For the specific investigation of the Gaussian fluctuations in 
BEC regime of the 2D crossover with analytical and numerical 
techniques see also Refs. \cite{sala2015,he2015}. 
{ Quite remarkably, our $T=0$ results with Gaussian fluctuations 
are in good agreement also with auxiliary-field 
path integral calculations \cite{shi2015} 
and diffusion Monte Carlo simulations \cite{galea2016}.} 

\begin{figure}[t]
\centerline{\epsfig{file=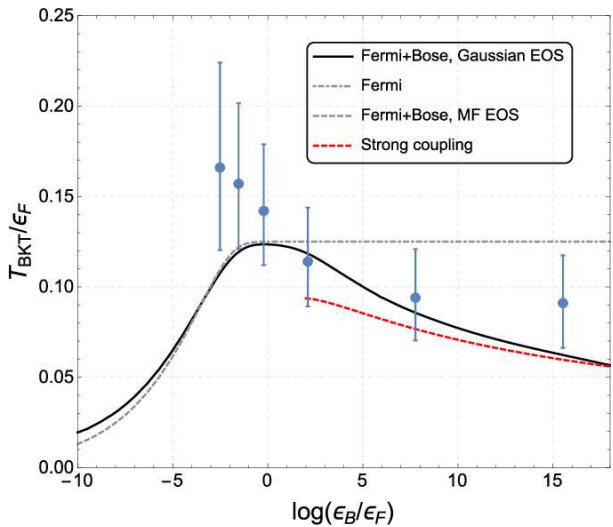,width=8cm,clip=}}
\caption{Our theoretical predictions \cite{bighin2016} for the 
Berezinskii-Kosterlitz-Thouless critical temperature $T_{BKT}$ 
compared to experimental observation \cite{murthy2015} 
(filled circles with error bars). Figure adapted from Ref. \cite{bighin2016}.} 
\label{fig4}
\end{figure}

We are now interested on the temperature dependence of 
superfluidy density $n_s(T)$ of the system.  
At the Gaussian level $n_s(T)$ depends only on fermionic 
single-particle excitations $E_{sp}(k)$ \cite{babaev1999}. 
Beyond the Gaussian level also bosonic collective 
excitations $E_{col}(q)$ contribute \cite{benfatto2004}. 
Thus, we assume the following Landau-type formula for the 
superfluid density \cite{bighin2016}
\beqa
n_s(T) &=& n - \beta \int
\frac{\mathrm{d}^2 k}{(2 \pi)^2} k^2 \frac{e^{\beta E_{sp}(k)}}
{(e^{\beta E_{sp}(k)} + 1)^2}
\nonumber
\\
&-& \frac{\beta}{2} \int \frac{\mathrm{d}^2 q}{(2 \pi)^2} q^2
\frac{e^{\beta E_{col}(q)}}{(e^{\beta E_{col}(q)} - 1)^2} \; . 
\label{trenta}
\eeqa
This bare superfluid density can be renormalized by using 
the flow equations (\ref{rg}) 
of Kosterlitz-Thouless-Nelson, which take into account 
the effect of quantized vortices and anti-vortices 
\cite{kos1972,nelson1977}, { by using Eq. (\ref{trenta}) 
as initial condition.}   
In Fig. \ref{fig4} we plot the BKT critical temparature obtained 
by using the Nelson-Kosterlitz criterion \cite{nelson1977}: 
\beq 
T_{BKT} = {\pi\over 8m} n_s(T_{BKT}) \; . 
\eeq
The figure clearly shows that the mean-field 
prediction (dashed line) is meaningful only in the deep 
BCS regime of the 2D crossover. Instead our beyond-mean-field results 
(solid line) which include Gaussian fluctuations, 
are in reasonable good agreement 
with the available experimental data \cite{murthy2015} (filled circles). 
In Ref. \cite{bighin2018,bighin2017} we have found that $T_{BKT}$ 
derived with the Nelson-Kosterlitz criterion slighlty overestimates 
the critical temperature calculated by solving the 
renormalization group equations (\ref{rg}). 

Having the equation of state and the superfluid density at finite temperature, 
we can calculate the first sound velocity $u_1$ and the 
second sound velocity $u_2$ in the 2D BCS-BEC crossover 
by using Eq. (\ref{boh}). 
We also analyze the amplitudes modes $W_1$and $W_2$ of the response to a 
density perturbation \cite{ozawa2014}, i.e. 
\begin{equation}
\delta n(x,t) = W_1 \delta n_1(x\pm u_1 t) + W_2\delta n_2(x\pm u_2 t)
\label{propagating perturbation}
\end{equation}
where
\begin{equation}
\frac{W_1}{W_1 + W_2}  = \frac{(u_1^2 - c_{20}^2)\, u_2^2}{(u_1^2 - u_2^2)
\, c^2_{20}} 
\label{w1}
\end{equation}
and
\begin{equation}
\frac{W_2}{W_1 + W_2} = \frac{(c_{20}^2 - u_2^2)\, u_1^2}{(u_1^2 - u_2^2)
\, c^2_{20}} \;.
\label{w2} 
\end{equation}

\begin{figure}[t]
\centerline{\epsfig{file=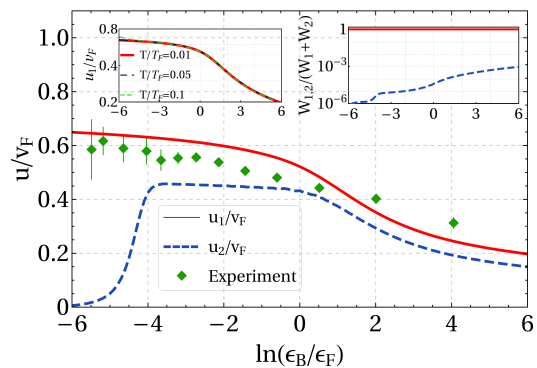,width=7cm,clip=}}
\caption{First sound velocity $u_1$ (red solid line) and second 
sound velocity $u_2$ (blue dashed line) 
along the BCS-BEC crossover, at temperature $T/T_F = 0.01$, with 
$T_F = \epsilon_F$ and $v_F = \sqrt{2\epsilon_F/m}$. 
Green diamonds: recent measurements of the first sound [M. Bohlen {\it et al.}
Phys. Rev. Lett. \textbf{124}, 240403 (2020).] 
Right inset: relative contribution to the density response 
of $u_1$ (red solid line) and $u_2$ (blue dashed line). 
Figure adapted from Ref. \cite{tononi2021}.}
\label{fig5}
\end{figure} 

In Fig. \ref{fig5} we show that our theoretical determination 
of the sound velocities and density responses (insets) as a function 
of the interaction strength (actually the logarithm of the 
adimensional binding energy $\epsilon_B/\epsilon_F$) at quite 
low temperature $T$. The comparison with the experimental 
measurements of the first sound velocity (filled diamonds) suggest that 
our theoretical framework is quite good. It is important to stress 
that in the BCS regime the speed of second sound is rapidly going to zero. 

\begin{widetext}

\begin{figure}[t]
\centerline{\epsfig{file=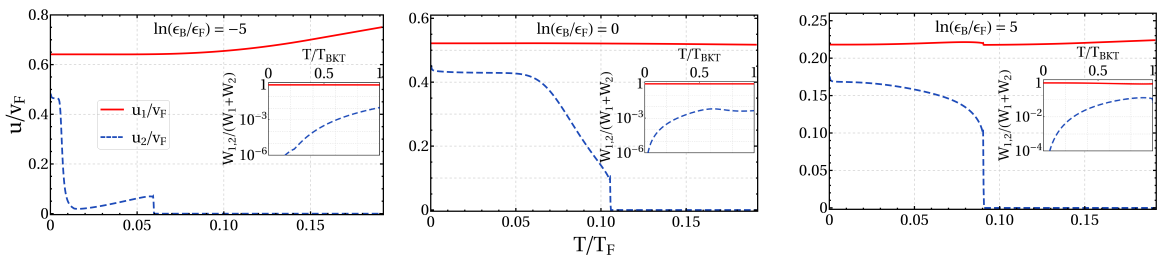,width=16cm,clip=}}
\caption{Adimensional first sound velocity $u_1/v_F$ (red solid line) 
and adimensional second sound velocity $u_2/v_F$ (blue dashed line) 
plotted in terms of the rescaled temperature $T/T_F$, for 
three different values of the crossover parameter: 
$\ln(\epsilon_B/\epsilon_F)=-5$ (BCS regime), 
$\ln(\epsilon_B/\epsilon_F)=0$ (unitary regime),
and $\ln(\epsilon_B/\epsilon_F)=5$ (BEC regime). 
Insets: relative contribution to the density responses
$W_{1,2}/(W_1+W_2)$ of $u_{1}$ and $u_2$. 
Figure adapted from Ref. \cite{tononi2021}.}
\label{fig6}
\end{figure} 

\end{widetext}

In Fig. \ref{fig6} we report instead the sound velocities as a function 
of the temperature $T$ for three values of the interaction strength 
(the three panels correspond to increasing values of the adimensional 
binding energy $\epsilon_B/\epsilon_F$). The density responses shown 
in the insets strongly suggest that a mixing between the first sound 
and second sound occurs only in the finite-temperature BEC regime. 
Notice that, taking int account Eqs. (\ref{c10}), (\ref{c20}), (\ref{cT}), 
(\ref{boh}), (\ref{w1}) and (\ref{w2}), the presence of mixing 
means that the adiabatic sound velocity 
$c_{10}$ is quite different with respect to the isothermal sound 
velocity $c_T$. Conversely, if $c_{10}\simeq c_T$ then $u_1\simeq c_{10}$, 
$u_{2}\simeq c_{20}$ and consequenty $W_{2}\simeq 0$. 
Just for comparison, for the 3D unitary Fermi gas we have recently 
shown \cite{bighin2022} that, contrary to 3D liquid helium, near the critical
temperature the mixing of first and second sound is quite strong. 

\section{Conclusions}

We have shown that first and second sound of bosonic and fermionic 
superfluids can be derived adopting the Landau's two-fluid theory 
which requires the equation of state and the superfluid fraction 
of the system under investigation. In the case of a 2D 
weakly-interacting Bose gas, we have found that the comparison 
of our theory with recent measurement near $T_{BKT}$ is quite good. 
In the BCS-BEC crossover of the 2D Fermi gas, we have proved that 
to obtain a good agreement with experimental data for the equation of state, 
the critical temperature $T_{BKT}$, and the sound modes, 
both fermionic single-particle excitations and bosonic collective 
excitations are needed. In conclusion, it is important to stress 
that all the results discussed here are valid in the collisional regime, 
where $\omega \tau\ll 1$ with $\omega$ the frequency of the sound 
mode and $\tau$ the collision time of quasi-particles. However, 
in the collisionless regime ($\omega \tau \gg 1$) 
the role of superfluidity in 2D systems of ultracold atoms 
is not yet fully clarified. 
In Ref. \cite{sattin2021} we have found that the experimental results 
of sound and sound damping in a two-dimensional collisionless 
Bose gas of $^{87}$Rb atoms \cite{dalibard2018} are better reproduced, 
below the critical temperature of the superfluid-to-normal phase 
transition, by the Andreev-Khalatnikov 
equations of a collisionless superfluid with respect to the finding of the 
Vlasov-Landau equation \cite{ota2018,cappellaro2018}. 
Finally, for the sake of completeness, we suggest to read the very recent  
review paper of Hu, Yao, and Liu \cite{hu2022} 
which contains a detailed historical account of the second sound 
in ultracold atoms. 

\vskip 0.3cm 
 
{LS thanks Herwig Ott and Sandro Wimberger for their kind 
invitation to the International Workshop 
``Quantum Transport with ultracold atoms'' (2022). 
KF acknowledges Fondazione CARIPARO for a PhD fellowship. 
AT acknowledges support from ANR Grant Droplets No. ANR-19-CE30-0003-02. 
LS and KF are partially supported by the BIRD Project 
"Ultracold atoms in curved geometries" of the University of Padova.}


\vskip 0.3cm

\begin{thebibliography}{99}

\bibitem{chris2021} P. Christodoulou, M. Galka, N. Dogra, R. Lopes, 
J. Schmitt, and Z. Hadzibabic, Observation of first and second sound 
in a BKT superfluid, Nature {\bf 594}, 191 (2021). 

\bibitem{furutani2021} K. Furutani, A. Tononi, and L. Salasnich, 
Sound modes in collisional superfluid Bose gases, New J. Phys. {\bf 23}, 
043043 (2021). 

\bibitem{bohlen2020} M. Bohlen, L. Sobirey, N. Luick, H. Biss, T. Enss, 
T. Lompe, and H. Moritz, Sound Propagation and Quantum-Limited Damping 
in a Two-Dimensional Fermi Gas, Phys Rev. Lett. {\bf 24}, 240403 (2020). 

\bibitem{tononi2021} A. Tononi, A. Cappellaro, G. Bighin, and L. Salasnich, 
Propagation of first and second sound in a two-dimensional Fermi superfluid, 
Phys. Rev. A {\bf 103}, L061303 (2021). 

\bibitem{landau1941} L.D. Landau, The theory of superfuidity of helium II, 
J. Phys. (USSR) {\bf 5}, 71 (1941). 

\bibitem{sala-power} L. Salasnich, Power spectrum and diffusion 
of the Amari neural field, Symmetry {\bf 11}, 134 (2019). 

\bibitem{gross1961} E. P. Gross, Structure of a quantized vortex 
in boson systems, Nuovo Cimento {\bf 20}, 454 (1961). 

\bibitem{pitaevskii1961} L. P. Pitaevskii, Vortex lines 
in an imperfect Bose gas, Sov. Phys. JETP {\bf 13}, 451 (1961). 

\bibitem{amari1977} S. Amari, Biol. Cyber. {\bf 27}, 77 (1977). 

\bibitem{sala2016} L. Salasnich and F. Toigo, Zero-point energy of 
ultracold atoms, Phys. Rep. {\bf 640}, 1 (2016). 

\bibitem{ber1972} V.L. Berezinskii, Destruction of Long-range Order 
in One-dimensional and Two-dimensional Systems Possessing a Continuous 
Symmetry Group. II. Quantum Systems, 
{\it Sov. Phys. JETP} {\bf 34}, 610 (1972).

\bibitem{kos1972} J.M. Kosterlitz and D.J. Thouless, 
Long range order and metastability in two dimensional solids and superfluids. 
(Application of dislocation theory), J. Phys. C {\bf 5}, L124 (1972).

\bibitem{nelson1977} D.R. Nelson and J.M. Kosterlitz, 
Universal Jump in the Superfluid Density of Two-Dimensional Superfluids, 
Phys. Rev. Lett. {\bf 39}, 1201 (1977). 

\bibitem{regal2004} C.A. Regal, M. Greiner, and D. S. Jin, 
Observation of Resonance Condensation of Fermionic Atom Pairs, 
Phys. Rev. Lett. {\bf 92}, 040403 (2004). 

\bibitem{zwierlein2004} M.W. Zwierlein, C. A. Stan, C. H. Schunck, 
S. M. F. Raupach, A. J. Kerman, and W. Ketterle, 
Condensation of Pairs of Fermionic Atoms near a Feshbach Resonance, 
Phys. Rev. Lett. {\bf 92}, 120403 (2004). 

\bibitem{kinast2004} J. Kinast, J. Kinast, S. L. Hemmer, 
M. E. Gehm, A. Turlapov, and J. E. Thomas, 
Evidence for Superfluidity in a Resonantly Interacting Fermi Gas, 
Phys. Rev. Lett. {\bf 92}, 150402 (2004). 

\bibitem{sala2015} L. Salasnich and F. Toigo, 
Composite bosons in the two-dimensional BCS-BEC crossover from 
Gaussian fluctuations, Phys. Rev. A {\bf 91}, 011604(R) (2015). 

\bibitem{he2015} L. He, H. Lu, G. Cao, H. Hu and X.-J. Liu, 
Quantum fluctuations in the BCS-BEC crossover of two-dimensional 
Fermi gases, Phys. Rev. A {\bf 92}, 023620 (2015).

\bibitem{makhalov2014} V. Makhalov, K. Martiyanov, and A. Turlapov, 
Ground-State Pressure of Quasi-2D Fermi and Bose Gases, 
Phys. Rev. Lett. {\bf 112}, 045301 (2014). 

{ \bibitem{shi2015} H. Shi, S. Chiesa, and S. Zhang, 
Ground-state properties of strongly interacting Fermi gases 
in two dimensions, Phys. Rev. A {\bf 92}, 033603 (2015).} 

{ \bibitem{galea2016} A. Galea, H. Dawkins, S. Gandolfi, and A. Gezerlis, 
Diffusion Monte Carlo study of strongly interacting two-dimensional 
Fermi gases, Phys. Rev. A {\bf 93}, 023602 (2016).} 

\bibitem{ries2015} M.G. Ries, A.N. Wenz, G. Zurn, L. Bayha, I. Boettcher, 
D. Kedar, P.A. Murthy, M. Neidig, T. Lompe, and S. Jochim, 
Observation of Pair Condensation in the Quasi-2D BEC-BCS Crossover, 
Phys. Rev. Lett. {\bf 114}, 230401 (2015). 

\bibitem{mora2003} C. Mora and Y. Castin, 
Extension of Bogoliubov theory to quasicondensates, 
Phys. Rev. A {\bf 67}, 053615 (2003). 

\bibitem{randeria1989} M. Randeria, J-M. Duan, and L-Y. Shieh, 
Bound states, Cooper pairing, and Bose condensation in two dimensions, 
Phys. Rev. Lett. {\bf 62}, 981 (1989).

\bibitem{nagaosa1999} N. Nagaosa, Quantum Field Theory in Condensed Matter 
(Springer, 1999). 

\bibitem{altland2006} A. Altland and B. Simons, Condensed Matter 
Field Theory (Cambridge Univ. Press, 2006). 

\bibitem{diener2008} R.B. Diener, R. Sensarma, 
Quantum fluctuations in the superfluid state of the BCS-BEC crossover, 
M. Randeria, Phys. Rev. A {\bf 77}, 023626 (2008).

\bibitem{taylor2006} E. Taylor, A. Griffin, N. Fukushima, 
Y. Ohashi, Pairing fluctuations and the superfluid density through the 
BCS-BEC crossover, Phys. Rev. A {\bf 74}, 063626 (2006).

\bibitem{bighin2016} G Bighin and L Salasnich, Finite-temperature 
quantum fluctuations in two-dimensional Fermi superfluids, 
Physical Review B {\bf 93}, 014519 (2016). 

\bibitem{bighin2018} G Bighin and L Salasnich, 
Renormalization of the superfluid density in the two-dimensional 
BCS-BEC crossover, Int. J. Mod. Phys. B {\bf 32}, 1840022 (2018). 

\bibitem{babaev1999} E. Babaev and H.K. Kleinert, 
Nonperturbative XY-model approach to strong coupling superconductivity 
in two and three dimensions, Phys. Rev. B {\bf 59}, 12083 (1999). 

\bibitem{benfatto2004} L. Benfatto, A. Toschi, and S. Caprara, 
Low-energy phase-only action in a superconductor: 
A comparison with the XY model, Phys. Rev. B {\bf 69}, 184510 (2004).

\bibitem{murthy2015} P.A. Murthy, I. Boettcher, L. Bayha, M. Holzmann, 
D. Kedar, M. Neidig, M.G. Ries, A.N. Wenz, G. Zurn, and S. Jochim, 
Observation of the Berezinskii-Kosterlitz-Thouless Phase Transition 
in an Ultracold Fermi Gas, Phys. Rev. Lett. {\bf 115}, 010401 (2015).

\bibitem{bighin2017} G Bighin and L Salasnich, 
Vortices and antivortices in two-dimensional ultracold Fermi gases, 
Sci. Rep. {\bf 7}, 45702 (2017). 

\bibitem{ozawa2014} T. Ozawa and S. Stringari, 
Discontinuities in the First and Second Sound Velocities at the 
Berezinskii-Kosterlitz-Thouless Transition, 
Phys. Rev. Lett. {\bf 112}, 025302 (2014). 

\bibitem{bighin2022} G. Bighin, A. Cappellaro, and L. Salasnich, 
Unitary Fermi superfluid near the critical temperature: 
thermodynamics and sound modes from elementary excitations, 
Phys. Rev. A {\bf 105}, 063329 (2022). 

\bibitem{sattin2021} F. Sattin and L. Salasnich, 
Collisionless sound of bosonic superfluids in lower dimensions, 
Phys. Rev. A {\bf 103}, 043324 (2021). 

\bibitem{dalibard2018} J.L. Ville, R. Saint-Jalm, E. Le Cerf, M. Aidelsburger,
S. Nascimbene, J. Dalibard, and J. Beugnon, 
Sound Propagation in a Uniform Superfluid Two-Dimensional Bose Gas, 
Phys. Rev. Lett. {\bf 121}, 145301 (2018).

\bibitem{ota2018} M. Ota, F. Larcher, F. Dalfovo, L. Pitaevskii, N.P.
Proukakis, and S. Stringari, 
Collisionless Sound in a Uniform Two-Dimensional Bose Gas, 
Phys. Rev. Lett. {\bf 121}, 145302 (2018).

\bibitem{cappellaro2018} A. Cappellaro, F. Toigo, and L. Salasnich, 
Collisionless dynamics in two-dimensional bosonic gases, 
Phys. Rev. A {\bf 98}, 043605 (2018).

\bibitem{hu2022} H. Hu, X.-C. Yao, X.-J. Liu, 
Second sound with ultracold atoms: A brief historical account, 
e-preprint arXiv:2206.05914. 

\end{thebibliography}
\end{document}